\documentclass[final,3p,times,twocolumn]{elsarticle}
\usepackage{amssymb}
\usepackage[latin1, ansinew]{inputenc}
\usepackage{graphicx}
\usepackage{bm}
\usepackage[T1]{fontenc}
\usepackage[dvips]{epsfig}
\usepackage{epsf}
\usepackage{color}

\begin{document}

\begin{frontmatter}

\title{MRI investigation of granular interface rheology \\using a new cylinder shear apparatus}

\author{Pascal Moucheront, François
Bertrand, Georg Koval~\footnotemark[1], Laurent Tocquer,
\\Stéphane Rodts, Jean-Noël Roux, Alain Corfdir and François
Chevoir~\footnotemark[2]}

\address{Université Paris-Est, UMR Navier (LCPC-ENPC-CNRS), Champs sur Marne, France}

\begin{abstract}
The rheology of granular materials near an interface is
investigated through proton magnetic resonance imaging. A new
cylinder shear apparatus has been inserted in the MRI device,
which allows the control of the radial confining pressure exerted
by the outer wall on the grains and the measurement of the torque
on the inner shearing cylinder. A multi-layer velocimetry sequence
has been developed for the simultaneous measurement of velocity
profiles in different sample zones, while the measurement of the
solid fraction profile is based on static imaging of the sample.
This study describes the influence of the roughness of the
shearing interface and of the transverse confining walls on the
granular interface rheology.
\end{abstract}

\begin{keyword}
velocimetry \sep granular material \sep annular shear \sep
interface \sep roughness

\end{keyword}

\end{frontmatter}

\footnotetext[1]{Now at the National Institute of Applied
Sciences (INSA), Strasbourg, France}

\footnotetext[2]{Corresponding author. Email address:
chevoir@lcpc.fr}

\section{Introduction} \label{sec:intro}

The interaction of a granular material with a solid interface is
of interest in various engineering problems, such as industrial
conducts~\cite{Nedderman92},
geotechnics~\cite{Schlosser78,Tubacanon95}, and in geophysical
situations, such as tectonophysics~\cite{Chambon06} or gravity
flows~\cite{Chevoir09}.

At the immediate vicinity of the shearing interface, a thin
granular layer, where the shear and dilation is localized, plays a
significant role in the stress transmission between the solid
interface and the bulk granular material. This rheology is
influenced by the roughness of the shearing
surface~\cite{DeJong03,Koval08,Oumarou05,Uesugi86a,Uesugi86b,
Uesugi88,Vardoulakis95}.

In this paper, we focus our attention on the annular
(\emph{Couette}) shear geometry, where the material is confined
between two cylinders and sheared by the rotation of the inner
rough one (see~\cite{Koval09a} for a recent review). This geometry
has been used to measure the rheological properties of granular
materials, both in two
dimensions~\cite{Howell99a,Latzel03,Miller96,Veje99} and three
dimensions~\cite{Bocquet02b,Chambon03,Dacruz04a,Dacruz02,Daniel07,
Losert01,Mueth03,Mueth00,Tardos98,Tardos03,Wang08}. However, the
visualization of the granular interface is usually limited to the
upper (free surface) or bottom layers (through a transparent glass
window) \cite{Chambon03,Koval08,Koval09b}. Following previous
magnetic resonance imaging investigation of granular
rheology~\cite{Fukushima99,Fukushima06,Kuperman96,Nakagawa93}:
flows in rotating
drum~\cite{Caprihan00,Ristow99,Seymour00,Yamane98}, vertical
chute~\cite{Chevoir01b}, annular shear
cell~\cite{Cheng06,Dacruz04a,Mueth00}, segregation and convection
under
vibration~\cite{Caprihan97,Ehrichs95,Hill97c,Knight96,Kuperman95,Metcalfe96,Yang00a,Yang02},
we have used MRI to measure the granular rheology (velocity and
solid fraction profiles) well inside the sample.

Sec.~\ref{setup} is devoted to the description of a new annular
shear cell, specially designed to be inserted in a MRI device.
Based on a geotechnical cylinder shear apparatus
\cite{Chambon03,Corfdir04,Dumitrescu05,Koval08,Koval09b,Lerat96,Lerat97},
its originality relies on the control of the radial confining
pressure exerted by the outer wall on the grains and on the
measurement of the torque on the inner shearing cylinder.

Sec.~\ref{methods} explains the multi-layer MRI velocimetry. MRI
velocity measurements were performed using a spin-warp / phase
encoding technique previously adapted from \cite{Hanlon98} and
used for routine liquid rheology in annular Couette cells (see
for instance \cite{Bonn08,Huang05,Jarny05,Ovarlez06,Ovarlez08}).
It was here further modified on purpose of quasi-simultaneous
assessment of different regions in the sample.

Sec.~\ref{results} describes the measurement of the velocity and
solid fraction profiles, as well as the shear stress on the
shearing wall, from which we deduce the influence of the roughness
of the shearing interface and of the transverse confining walls on
the granular interface rheology. Our study is restricted to the
quasi-static regime (small shear velocity and/or high confining
pressure).

Preliminary and complementary results are presented
in~\cite{Koval08}.

\section{Annular shear cell}\label{setup}

\begin{figure}[!htb]
\begin{center}
\includegraphics*[width=8cm]{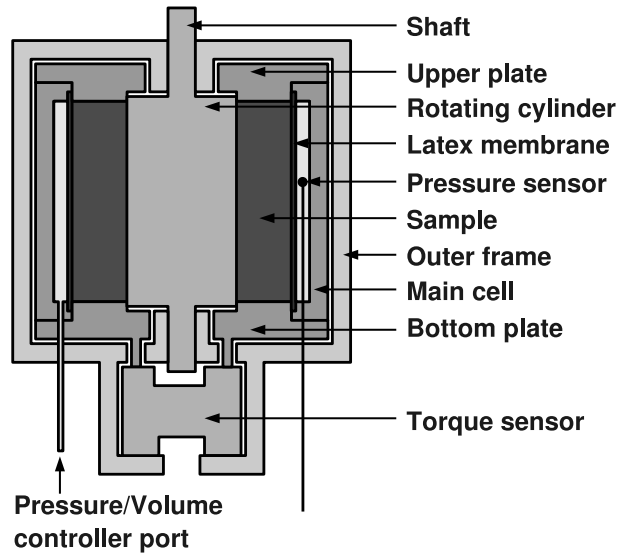}
\caption{\label{Fig1} \textit{Schematic view of the shear cell.}}
\end{center}
\end{figure}

The annular shear cell is inserted in the MRI device through an
external support. The figure \ref{Fig1} shows a schematic view of
the cell, which components are made of PMMA.

The sample has a hollow cylinder shape. The granular material is
confined between two fixed horizontal (bottom and upper) plates
(height $H=10~cm$), an internal rotating cylinder (radius
$R_{i}=3~cm$) and an external latex homemade dip molded membrane
(radius $R_{e}=6~cm$) filled with water. A pressure-volume
controller (GDS$^{\circledR}$) applies a radial confining pressure
$P$ to the sample, in the range $0-15~kPa$. An optical fiber
sensor (FOP MEMS $1000~kPa$ - Fiso Technologies$^{\circledR}$)
precisely measures this pressure close to the membrane.

The internal rotating cylinder is guided by two journal bearings
set in the external support, while the main cell (membrane, bottom
and upper plates) is only connected to the external support
through the torque sensor, which measures the whole torsion
effort.

An aluminum-alloy torque sensor was specially designed, based on
resistive strain gages, in the range $\pm 10$ Nm. This measurement
was not possible during MRI experiment, but when displacing the
cell bellow the MRI device. This prototype is a first step toward
the realization of a torque sensor working inside the MRI device.

Depending on the way upper and bottom plates are mounted, it is
possible to measure the whole torsion effort or only the fraction
transmitted to the lateral membrane (the difference between those
two measurements provides the torque transmitted by the horizontal
walls).

The cell is connected to the transmission axis of a rheometer
previously designed to be inserted within the MRI facility
\cite{Moller08,Ovarlez06,Raynaud02,Rodts04}, through a gearbox.
This two-stage, timing-belt and pulley system is placed close to
the cell and far from the motor. Its reduction factor of $10$
provides a rotation range $1/600\leq \Omega \leq 1/6$ RPS.

This configuration allows to place down the cell (out of the MRI
tunnel) during the sample preparation and torque measurement and
then move it up to the observation position.

The complete cell has a total diameter of $\approx 19.5~cm$ and a
total height, without gearbox, of $28~cm$, which fits inside the
RF coil.

\section{Multi-layer MRI velocimetry}\label{methods}

MRI experiments were carried out on a Bruker Biospec $24/80$ MRI
facility operating at $0.5~T$ ($21~MHz$ proton frequency). The MRI
magnet is a vertical superconducting prototype (Magnex Scientific,
Oxford), with a $40~cm$ bore. These characteristics are
particularly suited for the study of large and inhomogeneous
samples, exhibiting strong internal susceptibility contrasts. The
magnet is equipped with a birdcage RF coil (height: $20~cm$, inner
diameter: $20~cm$, hard $\pi/2$ pulse duration: $100\mu s$), and a
BGA26 shielded gradient system (Bruker), delivering a $0.05~T/m$
gradient strength with a rising time of $500\mu s$.

\begin{figure}[!htb]
\begin{center}
\includegraphics*[width=6.5cm]{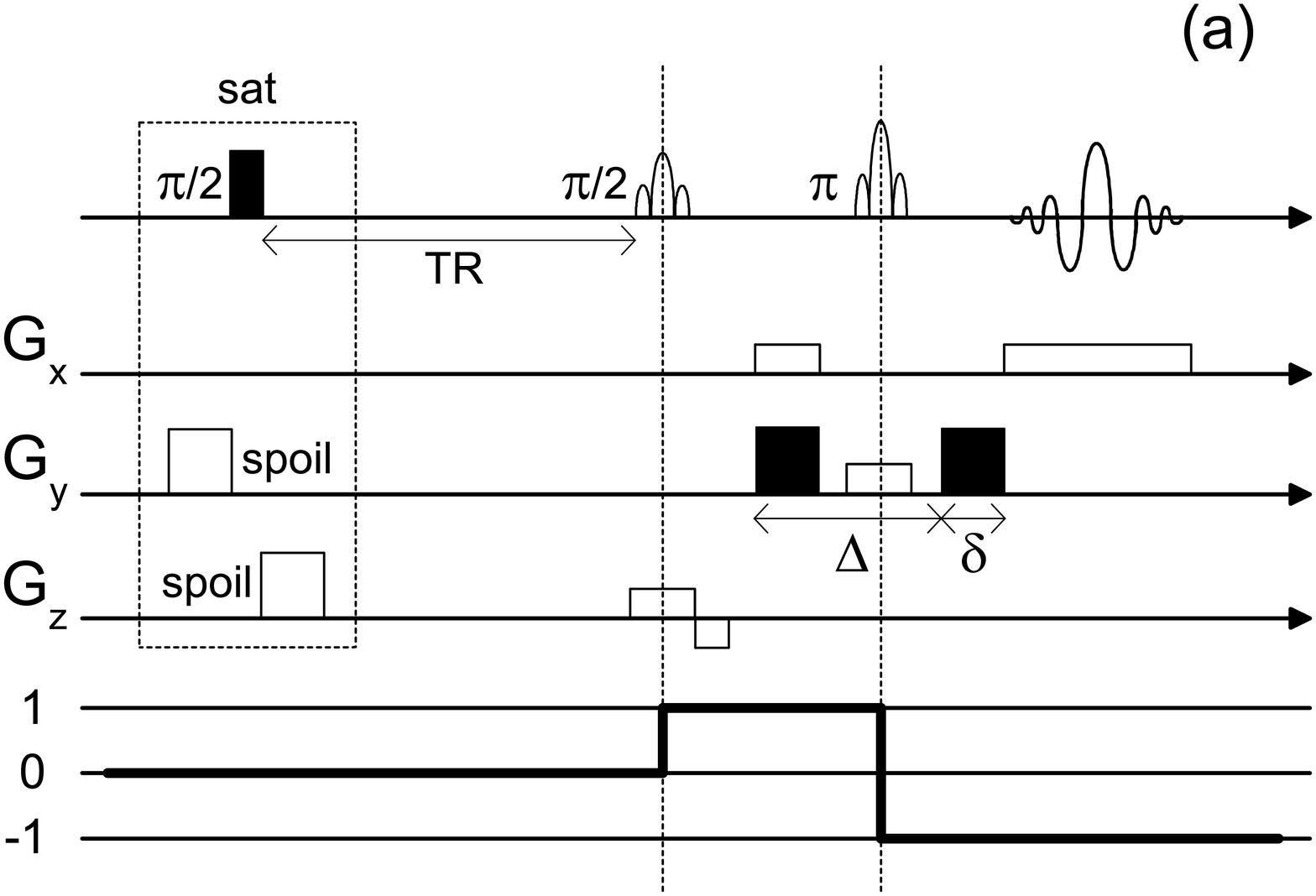}
\includegraphics*[width=6cm]{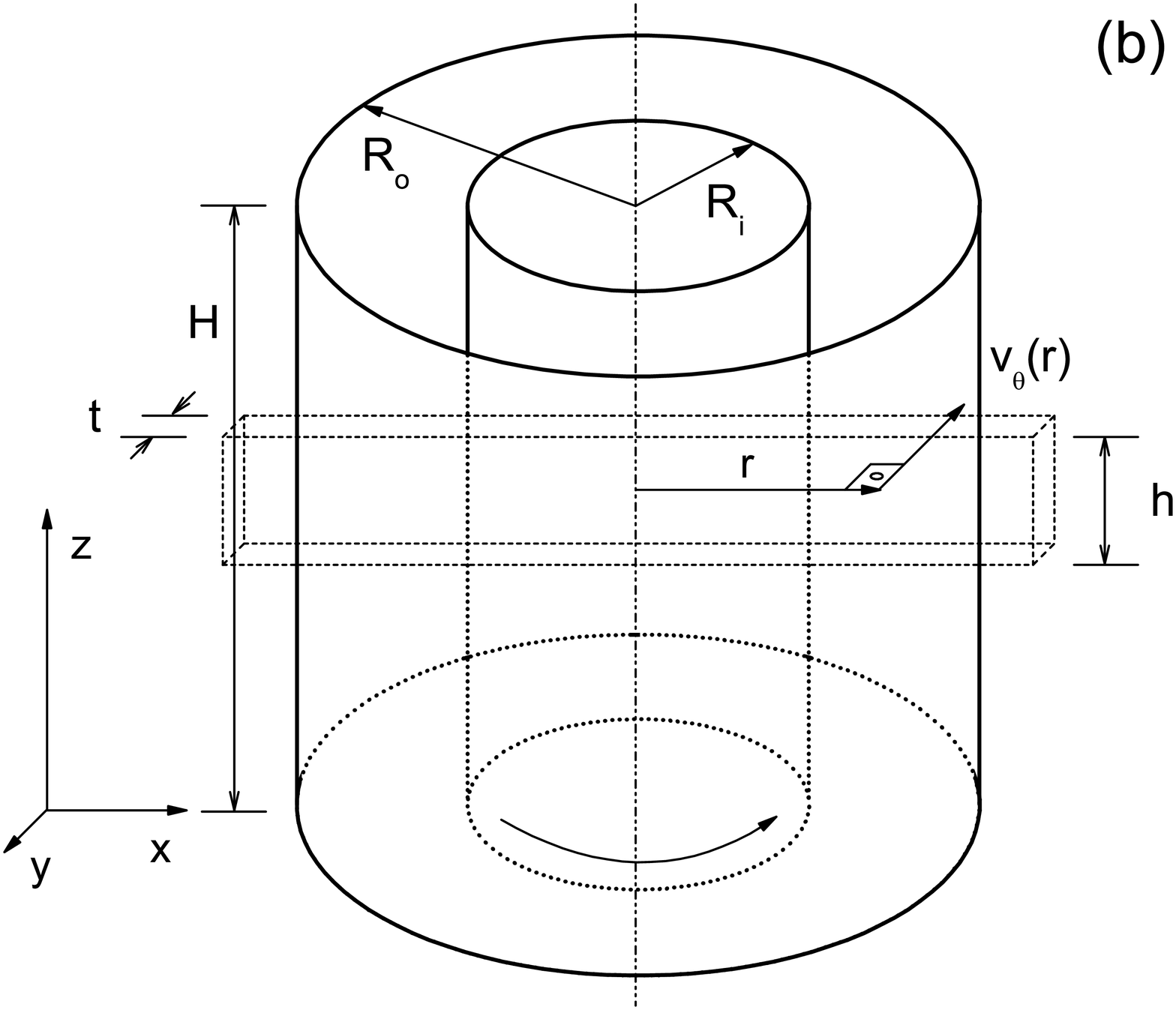}
\caption{\label{Fig2} \textit{(a) MRI sequence used for velocity
measurements, and its coherence pathway. Velocity encoding
gradients are shown in black. (b) Cell scheme and virtual beam
cut.}}
\end{center}
\end{figure}

MRI methodology for radial velocimetry inside the cell was that of
\cite{Hanlon98} as further modified by~\cite{Raynaud02}
and~\cite{Rodts04}. It is based on a two-pulse spin echo sequence
(Fig.~\ref{Fig2}a), in which the two pulses are made
space-selective in $z$ and $y$ direction respectively so as to
virtually cut a beam along one cell diameter (Fig.~\ref{Fig2}b). A
read-out gradient in $x$ directions permits to get after Fourier
transform of the signal a 1D magnetization profile along the beam.
An additional pair of gradient pulses in $y$ direction (in black)
induces on the magnetization profile an $x$ dependent phase shift
reading:

\begin{equation}
\varphi(x)=\gamma G \delta \Delta v_y(x),
\end{equation}

\noindent where $\gamma$ is the gyromagnetic ratio of proton, $G$
the strength of these 'velocity' pulses, $\delta$ and $\Delta$ are
timing components of the sequence as described in Fig.~\ref{Fig2},
and $v_y(x)$ is the $y$ velocity component along the selected
diameter. In order to get a velocity profile, one performs two MRI
measurements with respectively positive and negative velocity
gradients, and then compares in each pixel the phase of the two
magnetization profiles. When the thickness of the beam in $y$
direction is small enough as compared with the cell diameter, such
measurement may be regarded as a direct measurement of the
orthoradial velocity component $v_{\theta}$ versus the radial
coordinate $r$ (Fig.~\ref{Fig2}b).

In this sequence, only the phase of NMR signal bears relevant
information. It is then safe to use recycling times $TR$ equal or
even shorter than $T_1$ values. Indeed, an incomplete spin-lattice
relaxation between two successive sequences may only affect the
signal to noise ratio, but is not prone to introduce any bias,
provided that the magnetization be correctly cleaned from past
solicitations before each sequence. Such cleaning was here
performed with a saturation module at the beginning of the $TR$
delay.

Thanks to saturation, $TR$ may even be chosen so as to optimize
the signal to noise ratio of the experiment. Let's imagine that,
because of experimental constraints, one has a fixed delay
$T_{exp}$ to perform a velocity measurement. If some $TR$ delay is
used, then it will be possible to repeat the sequence
$N=T_{exp}/TR$ times. Granted that the available magnetization at
the beginning of each sequence scales as $1-\exp(-TR/T_1)$, and
that the improvement of signal to noise ratio when repeating
sequences scales as $\sqrt{N}$, the signal to noise ratio $SNR$ of
the whole measurement process scales as:

\begin{equation}
SNR(TR) \propto \left\{1-\exp\left( -\frac{TR}{T_1} \right)
\right\}\sqrt{\frac{T_{exp}}{TR}}.
\end{equation}

\noindent This quantity is maximum for:

\begin{equation}
TR_{opt}=1.26T_1,
\end{equation}

\noindent and fairly stays above 90\% of this maximum in the range
$0.57 T_1 \leq TR \leq 2.59 T_1$ . Saturation then turns out a
useful tool for signal optimization.

For the assessment of the velocity profiles at different height in
$z$ direction, sequences were repeated while systematically
shifting the vertical position of the beam. In order to study
unsteady granular flows, it was important to make these
multi-layer measurements as simultaneous as possible.

First of all, looking back at the NMR sequence, the measurement of
one beam at a given position clearly affects all the magnetization
of a larger region in the cell, composed of the sum of the two
slices selected by each RF pulse. This, together with the use of
the saturation module, prevents considering measurements at
different $z$ positions as independent: it is not possible to run
the sequence at one height immediately after running the sequence
at another height, even if the two associated beams are well
separated. Real simultaneous measurements of velocity in multiple
layers were then not possible.

\begin{figure}[!htb]
\begin{center}
\includegraphics*[width=7.5cm]{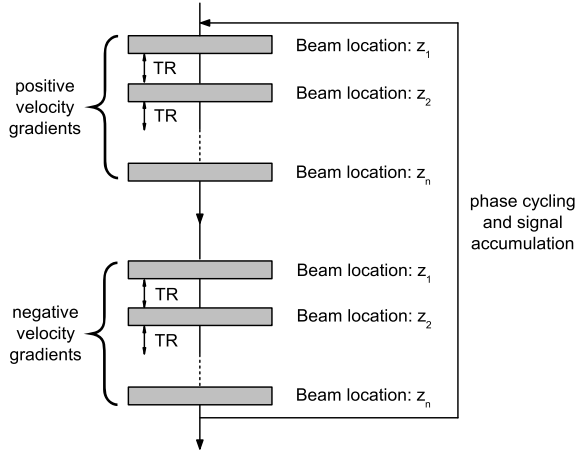}
\caption{\label{Fig3} \textit{Organization of MRI velocimetry
sequences for the quasi-simultaneous assessment of n layers in the
cell.}}
\end{center}
\end{figure}

We proceeded instead as follows (Fig~\ref{Fig3}). The whole
measurement consisted in two nested loops. In the first loop the
vertical location of the beam was shifted after each single
sequence, so as to go through all necessary beam positions. The
procedure was then just repeated with inverted velocity gradient
strength. This first loop was inserted in a second one consisting
in standard phase cycling and signal accumulation. One could get
this way velocity profiles at different height measured over
exactly the same period of time.

In the present work, the thickness $t$ of the beam in $y$
direction was $10~mm$, and the recycling time $TR$ was taken as
$100~ms$. The typical echo-time was about $20~ms$. Strong enough
velocity gradients were here sufficient to select the proper
coherence pathway. Our MRI system did not suffer any visible
acquisition quadrature defect, so that the phase cycling procedure
simply consisted here in a base-line correction, and could be
completed in two steps. The minimal value of the height $h$ of the
beams was $1.5~mm$. The maximum relevant value was the cell size.
Thinner layers are well adapted to study regions with a strong
vertical gradient of velocity, like the zone close to the
horizontal walls. Because of hardware constraints, the maximum
number of layers was $10$. The total experimental time was 13
minutes for three layers of $1.5~mm$ width, 104 minutes for five
layers of $0.4~mm$ width and 208 minutes for five layers of
$0.15~mm$ width.

\section{Experimental results}\label{results}

Three roughness levels of the rotating cylinder are tested: the
first one corresponds to a vertically corrugated surface
(triangular shape with thickness equal to depth of $1~mm$), the
second one to a glued seeds surface and the third one to sandpaper
(medium grain - 80). The horizontal walls (upper and bottom) are
originally smooth. To modify their roughness we have sticked
sandpaper and glass beads with the same diameter as the seeds (for
practical reasons the beads are glued to sandpaper and not
directly over the plate surface).

Mustard seeds are used as model granular material (mean diameter
$d=1.5~mm$, quite monodisperse, mass density of $1200~kg/m^3$).
The relaxation times are $T_1 \approx 100~ms$ and $T_2 \approx
40~ms$.

\begin{figure}[!htb]
\begin{center}
\includegraphics*[width=5cm]{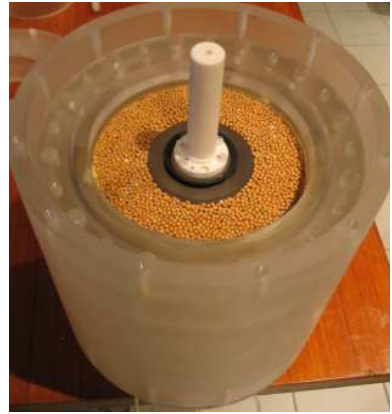}
\caption{\label{Fig4} \textit{(Color on line) Sample
preparation.}}
\end{center}
\end{figure}

The granular material is poured inside the cell with a funnel from
the top and then levelled off (see Fig.\ref{Fig4}). After closing
the cell, the sample is confined by the application of the radial
pressure through the membrane. Then, the material is pre-sheared
(at least $15$ rotations of the cylinder, corresponding to
$\approx 2800~mm$ of tangential displacement of the wall) to drive
it in a steady-state. Then data acquisition starts.

Tab.~\ref{Tab1} summarizes the parametric study (wall roughness,
confining pressure $P$ and velocity at the wall $V_{\theta}= 2 \pi
R_i \Omega$).

\begin{table}[!htb]
  \centering
\begin{tabular}{|c|c|c|c|}
  \hline
  \textbf{cylinder} & \textbf{plates} & \textbf{pressure$^1$ } & \textbf{velocity$^2$}  \\
  \hline
  corrugated             & smooth                       & $3.5-13.5$ & $0.314 $|3.14 \\
  \hline
  glued seeds       & smooth                       & $3.5-13.5$ & $0.314$|3.14 \\
                    & sandpaper             & 8.5                & 3.14                   \\
                    & glass beads     & 8.5                & 3.14                   \\
  \hline
  sandpaper   & smooth                       & $3.5-13.5$ & $0.314$|3.14 \\
  \hline
\end{tabular}
$^1$ pressure (kPa) ~~ $^2$ velocity (mm/s) \caption{\label{Tab1}
\textit{List of velocimetry experiments.}}
\end{table}

Internal stresses associated to gravity (of the order of 0,7 kPa)
can be considered to be dominated by the applied confining
pressures.

The maximum value of the inertial number \cite{Koval09a}, used to
qualify the granular flow regime, can be estimated around $3 \cdot
10^{-4}$, indicating that the granular material is in the
quasi-static regime.

We now discuss the radial profiles of velocity and solid fraction.
They are presented as a function of $r-R_i$, normalized by the
diameter $d$ of the grains.

\subsection{Velocity profiles}\label{velprofiles}

\begin{figure}[!htb]
\begin{center}
\includegraphics*[width=7cm]{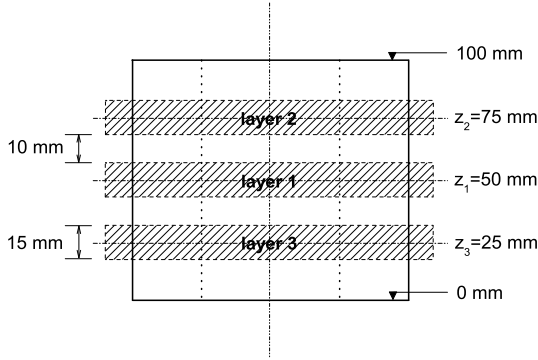}
\caption{\label{Fig5} \textit{Scheme of layers in the central
zone.}}
\end{center}
\end{figure}

We have first measured the profiles of the orthoradial velocity
$v_{\theta}(r)$, normalized by $V_{\theta}$, far from the
horizontal walls, comparing three layers $15~mm$ thick separated
by $10~mm$ (Fig.~\ref{Fig5}).

\begin{figure}[!htb]
\begin{center}
\includegraphics*[width=7cm]{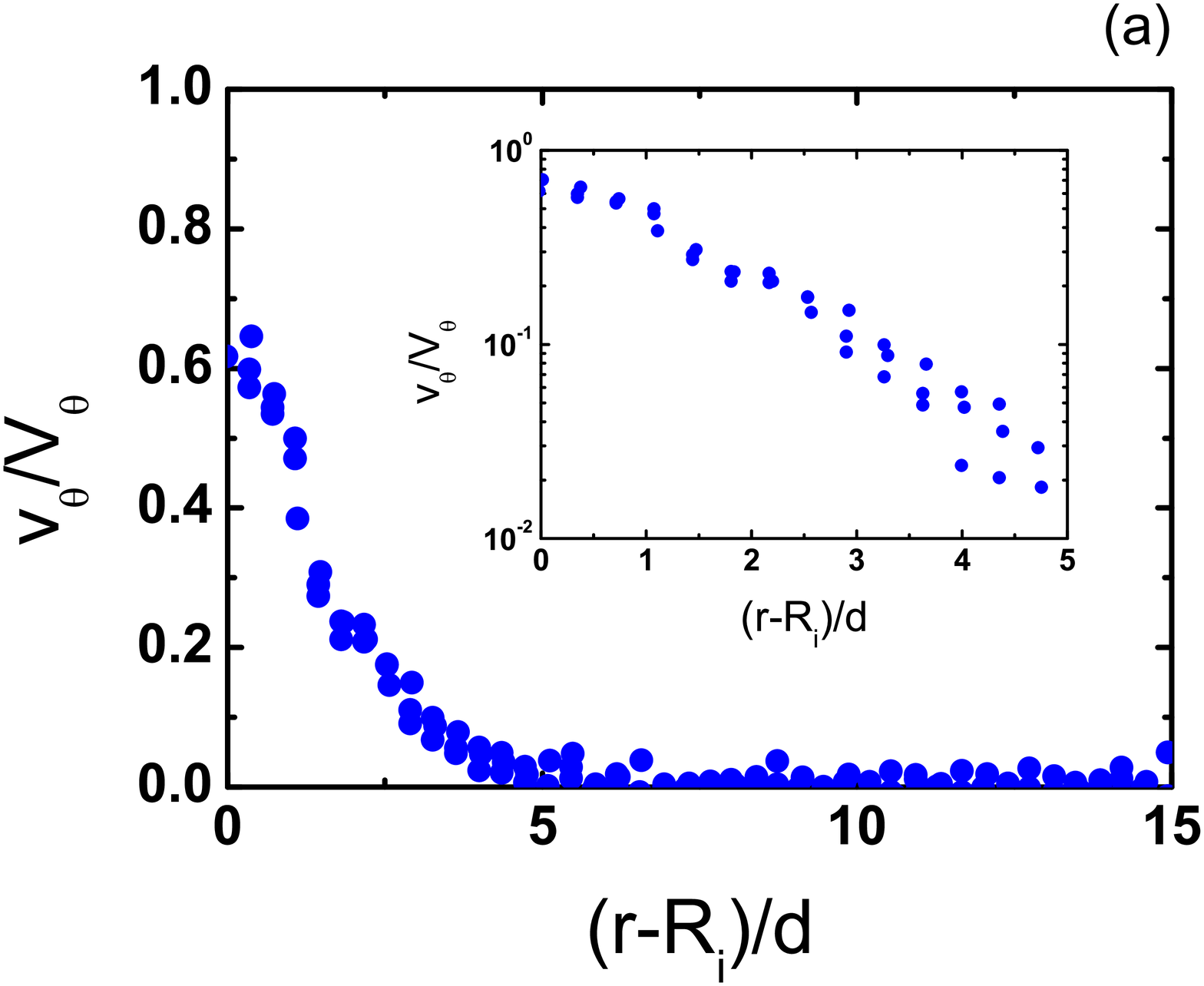}
\includegraphics*[width=7cm]{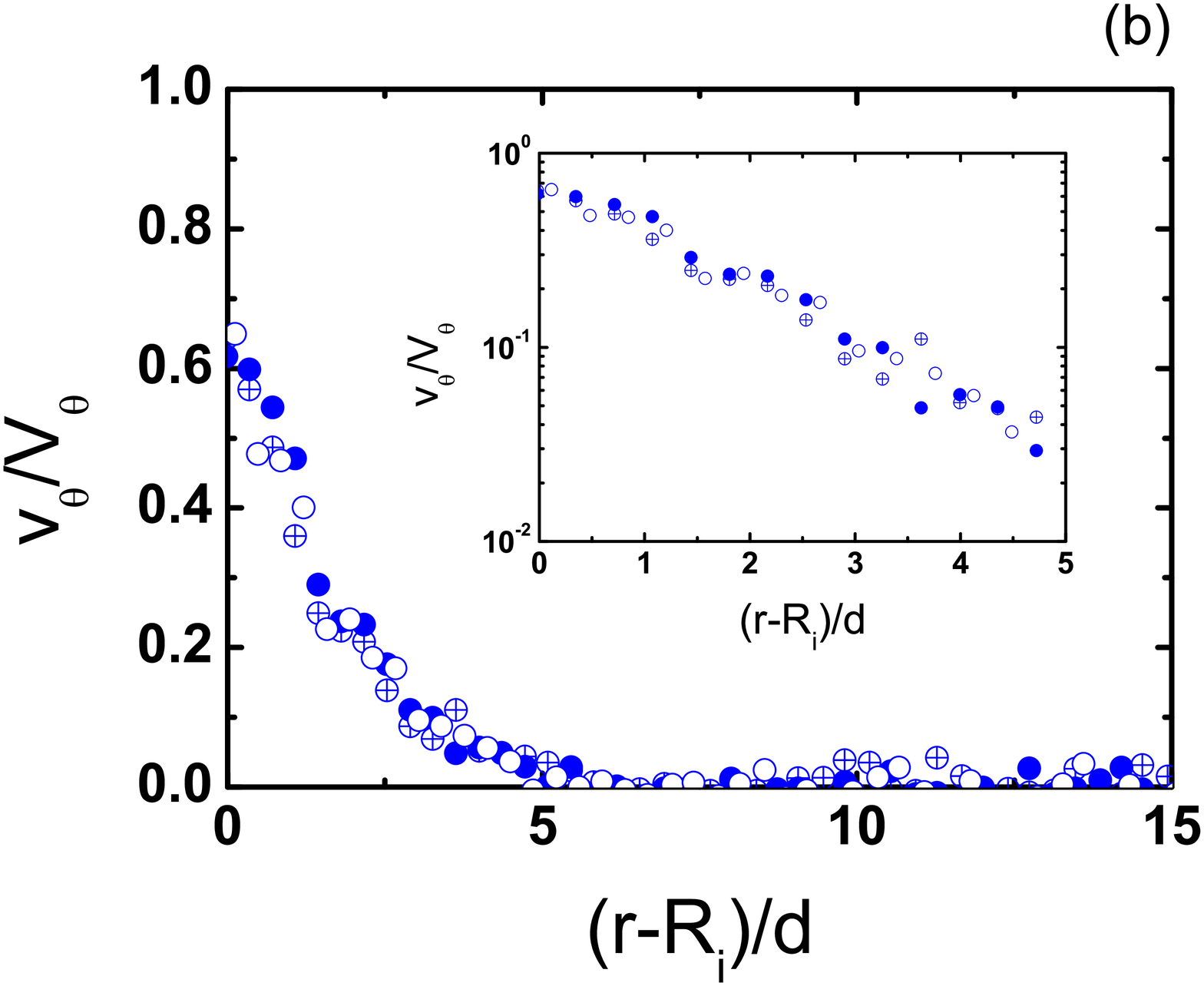}
\caption{\label{Fig6} \textit{(Color on line) Velocity profiles
(glued seeds cylinder). (a) Smooth horizontal walls (layers $1$,
$2$ and $3$). (b) Influence of the roughness of the horizontal
walls (layer $1$): ($\textcolor[rgb]{0.00,0.00,1.00}{\bullet}$)
smooth, ($\textcolor[rgb]{0.00,0.00,1.00}{\oplus}$) sandpaper,
($\textcolor[rgb]{0.00,0.00,1.00}{\circ}$) glued glass beads. In
inserts, the region closer to the shearing cylinder in
semi-logarithmic scale.}}
\end{center}
\end{figure}

Fig.\ref{Fig6} shows that $v_{\theta}(R_i)  < V_{\theta}$, which
signifies that there is significant sliding at the wall.
Fig.\ref{Fig6} also shows that there is no influence of the
roughness of the horizontal walls in the central zone. An
approximately exponential decay of the orthoradial velocity
profiles $v_{\theta}(r)$ is measured, with a maximum value at the
shearing wall, consistently with previous
studies~\cite{Bocquet02b,Chambon03,
Dacruz04a,Howell99a,Koval09b,Koval09a,Latzel03,Mueth00}.
Qualitatively, this shear localization is explained by the strong
decrease of the shear stress away from the inner wall, while the
normal stress remains approximately constant. As expected in the
quasi-static regime, no influence of the velocity at the wall
$V_{\theta}$ nor of the confining pressure $P$ has been observed.

\subsubsection{Influence of the roughness of the
horizontal walls}

In a $2D$ system, the shear stress is entirely transmitted  from
the inner to the outer wall by the granular
material~\cite{Koval08,Koval09a}. In a $3D$ system, the granular
material interacts with the horizontal walls, to which part of the
torque is transmitted. This changes the stress distribution within
the sample, and may consequently affect the $3D$ velocity profile.
This influence of frictional lateral walls has been evidenced in
granular surface flows between vertical walls~\cite{Jop05b},
especially when they are rough~\cite{Jop06}. We now wonder if such
effect is observed close to the horizontal walls.

\begin{figure}[!htb]
\begin{center}
\includegraphics*[width=7cm]{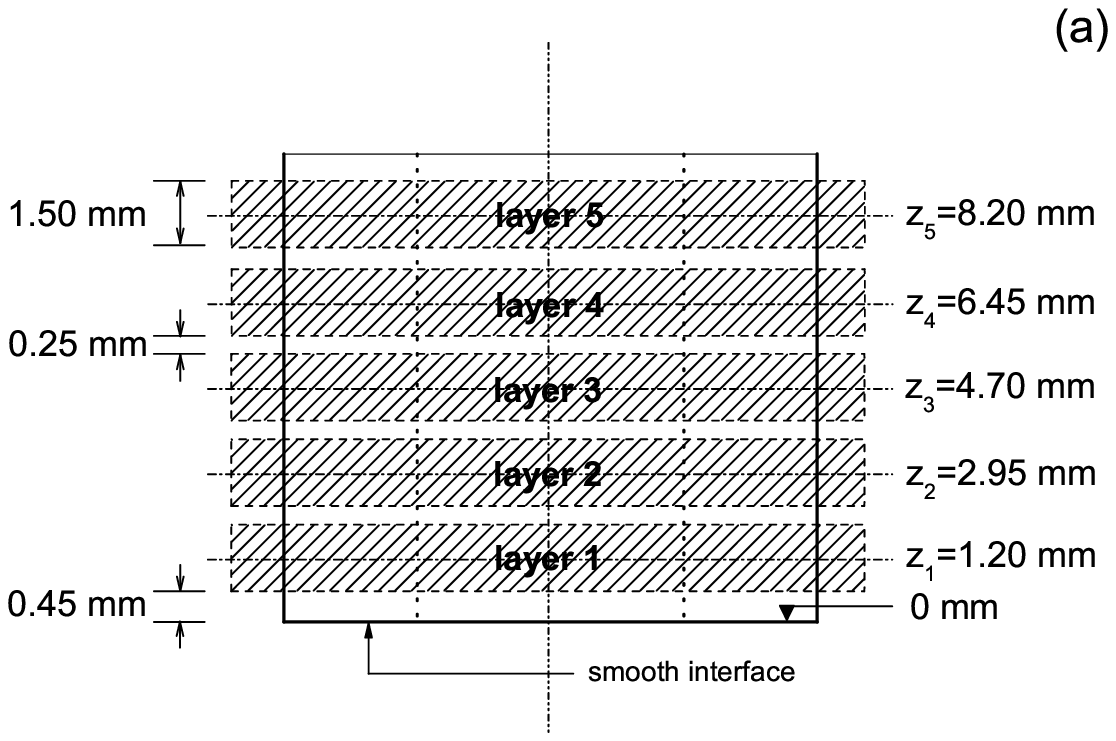}
\includegraphics*[width=7cm]{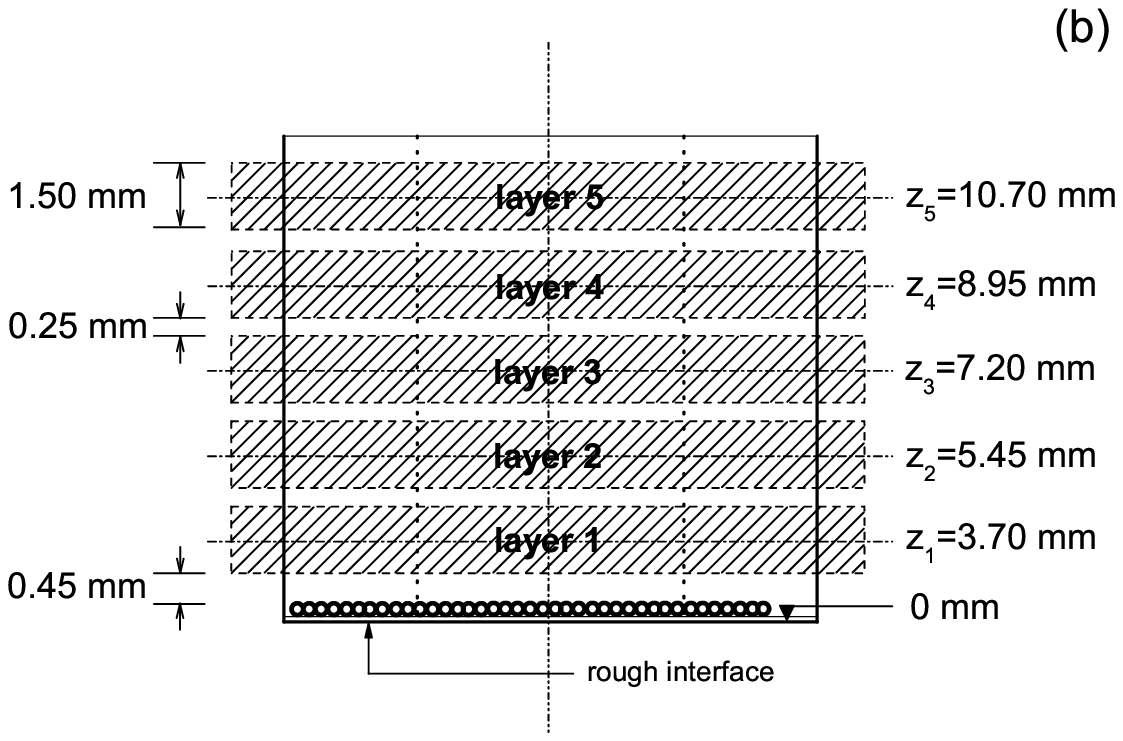}
\caption{\label{Fig7} \textit{Scheme of layers close to the
inferior plate.}}
\end{center}
\end{figure}

The analysis of the influence of the horizontal walls was based on
a comparison of the profiles of $5$ layers $1.5~mm$ thick
 separated by a distance of $1.75~mm$ (Fig.~\ref{Fig7}). The gap of
$0.45~mm$ between the horizontal wall and the first layer ensured
that only moving particles were considered in the measure. In the
rough case, a similar gap was taken from the glued glass beads
surface giving a total thickness of $2.5~mm$.

The horizontal roughness made of glass beads is not detectable by
MRI, which provided a correct evaluation of the velocity of the
grains in direct contact with this surface. The motion of the
grains could be clearly distinguished from the motion of the
cylinder, given the strong velocity gradient.

\begin{figure}[!htb]
\begin{center}
\includegraphics*[width=7cm]{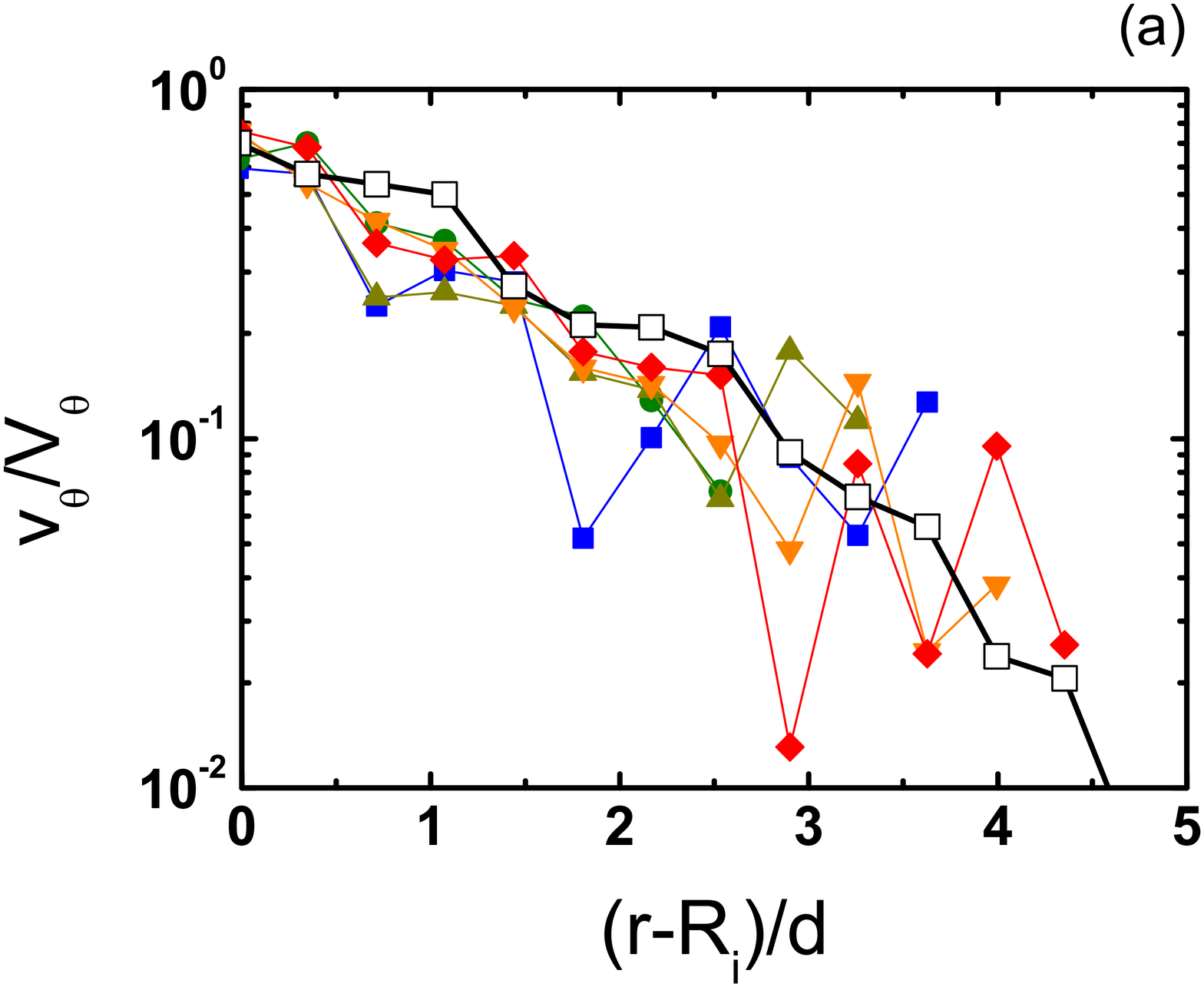}
\includegraphics*[width=7cm]{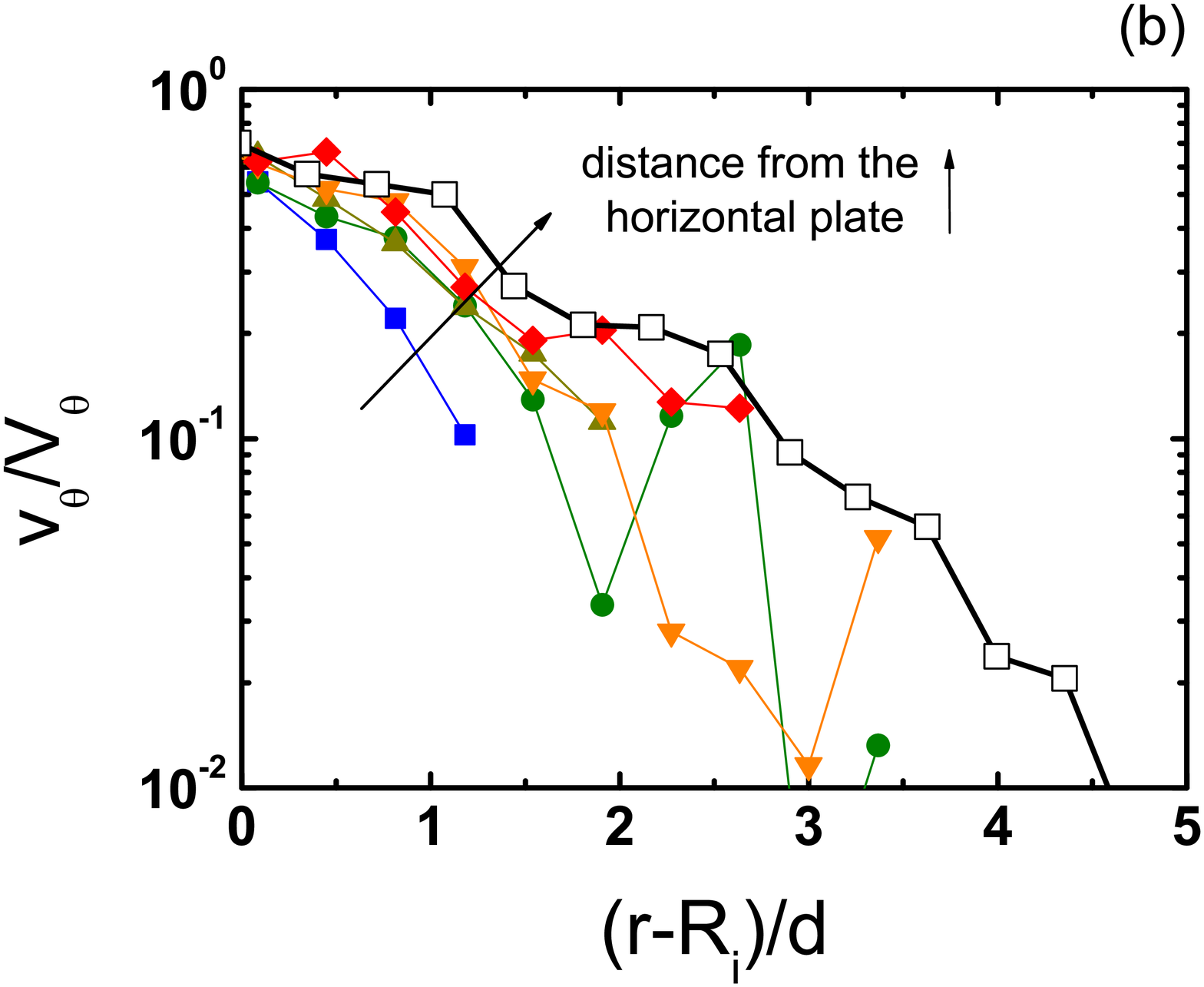}
\caption{\label{Fig8} \textit{(Color on line) Velocity profiles
near the bottom horizontal wall (semi-logarithmic scale, glued
seeds cylinder). (a) Smooth horizontal wall (layers described in
Fig.~\ref{Fig7}a), (b) horizontal wall with glued glass beads
(layers  described in Fig.~\ref{Fig7}b).
($\textcolor[rgb]{0.00,0.00,1.00}{\blacksquare}$) layer~$1$,
($\textcolor[rgb]{0.00,0.59,0.00}{\bullet}$) layer~$2$,
($\textcolor[rgb]{0.50,0.50,0.00}{\blacktriangle}$) layer~$3$,
($\textcolor[rgb]{1.00,0.50,0.25}{\blacktriangledown}$) layer~$4$,
($\textcolor[rgb]{0.98,0.00,0.00}{\blacklozenge}$) layer~$5$,
($\square$) central profile.}}
\end{center}
\end{figure}

Fig.~\ref{Fig8} shows a comparison between the profiles in the
central zone and near the horizontal wall. In the case of a smooth
horizontal wall, there is no significant influence. This result
has practical applications, validating the measure of
displacements of granular materials through transparent glass
walls~\cite{Chambon03,Koval08,Koval09b}. Conversely, in the case
of a rough horizontal wall, a significant decrease is observed
very close to the horizontal wall, particularly in the first
layer. This indicates the transmission of shear stress between the
granular material and the horizontal walls. However, the
perturbation remains very localized, in contrast with granular
surface flows~\cite{Jop06}. The fluctuations of the velocity
profiles in the first layers close to the horizontal wall is
explained by the small thickness of those layers, which strongly
decreases the signal to noise ratio.

\subsubsection{Influence of the cylinder roughness} \label{rugoVmu}

\begin{figure}[!htb]
\begin{center}
\includegraphics*[width=7cm]{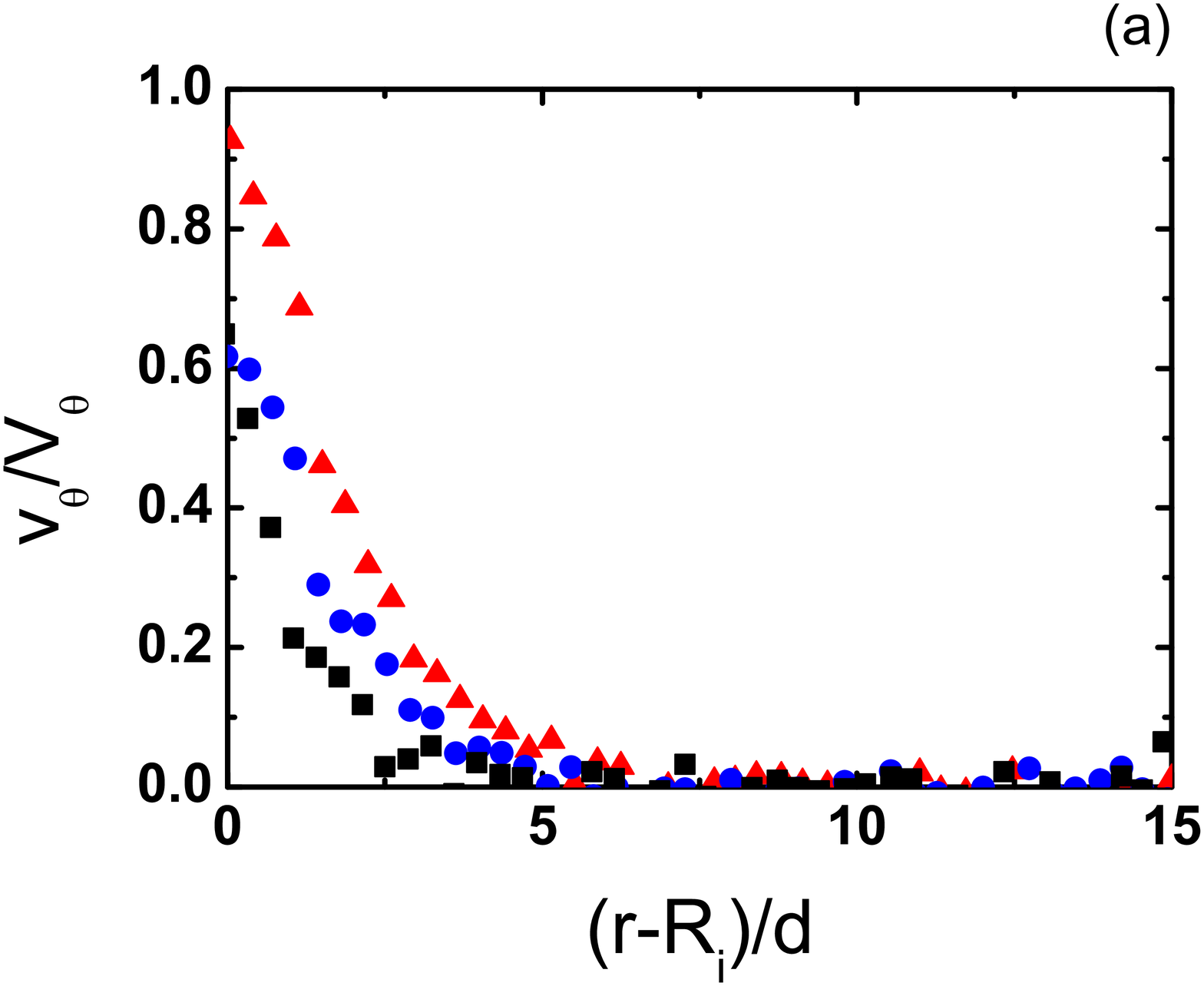}
\includegraphics*[width=7cm]{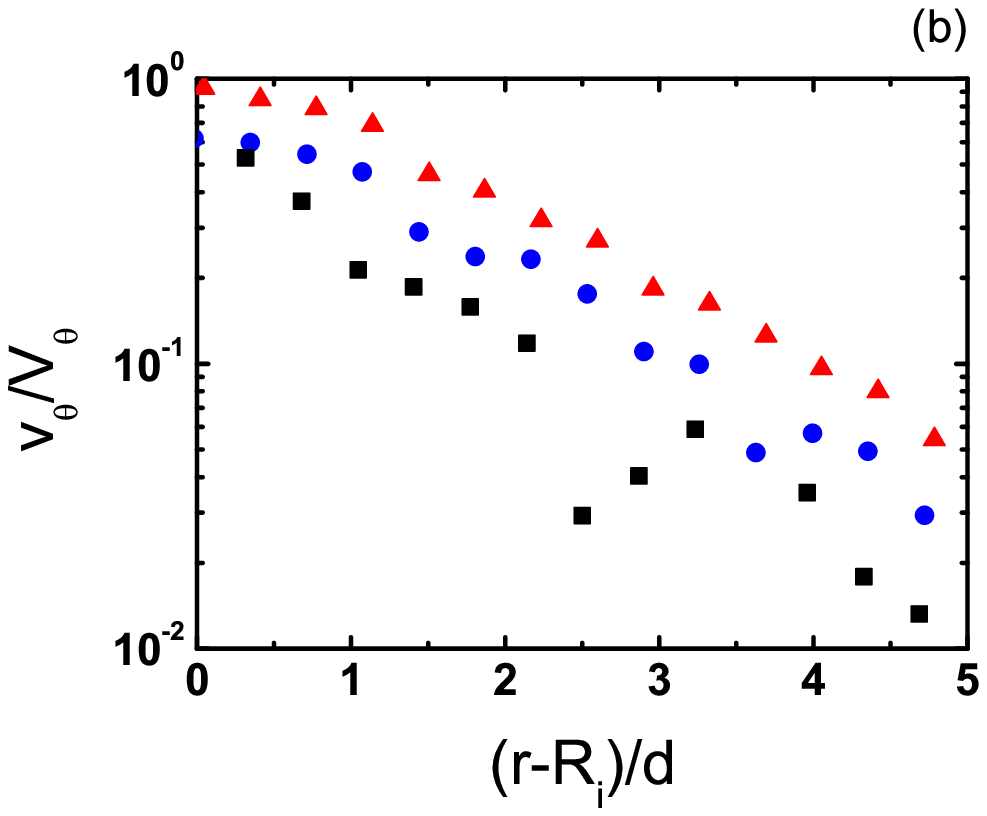}
\caption{\label{Fig9} \textit{(Color on line) Influence of the
cylinder roughness on the velocity profiles: ($\blacksquare$)
sandpaper, ($\textcolor[rgb]{0.00,0.00,1.00}{\bullet}$) glued
seeds and ($\textcolor[rgb]{0.98,0.00,0.00}{\blacktriangle}$)
corrugated. (a) Linear scale and (b) detail of the region near the
shearing cylinder in semi-logarithmic scale.}}
\end{center}
\end{figure}

Fig.~\ref{Fig9} shows that the cylinder roughness has a
significant influence on the velocity $v_{\theta}$, causing
offsets of the whole velocity profiles. Sliding, defined as the
ratio between the maximum value of the particle velocity and the
wall velocity $V_{\theta}$~\cite{Koval09b,Koval09a}, decreases
from sandpaper cylinder to the corrugated one, that is to say as
the roughness increases.

\subsection{Friction coefficient}

\begin{figure}[!htb]
\begin{center}
\includegraphics*[width=7.5cm]{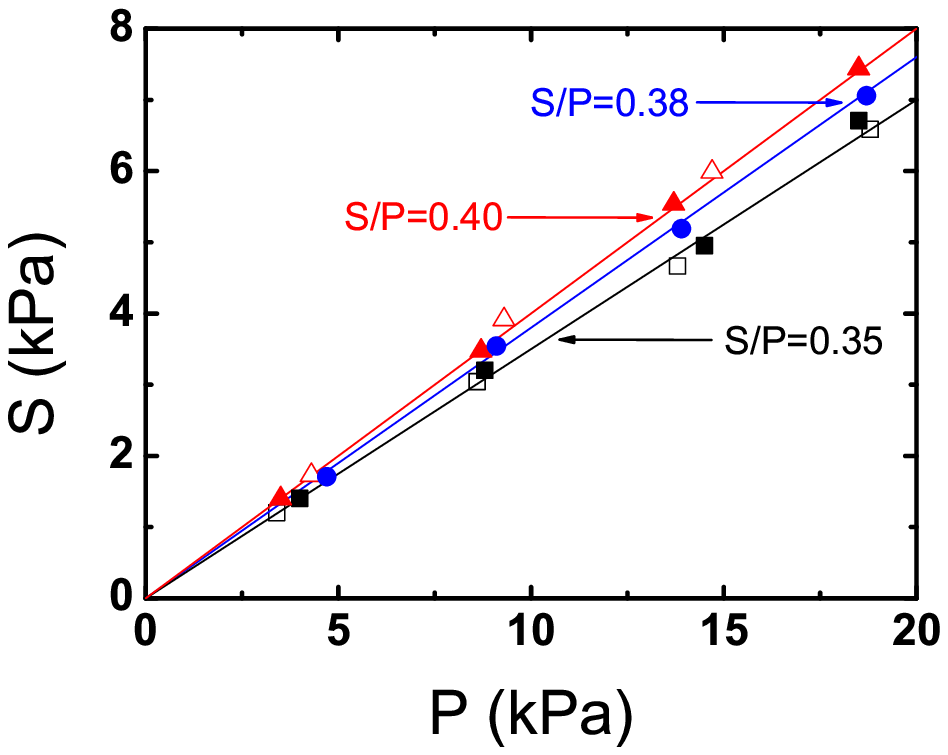}
\caption{\label{Fig10} \textit{(Color on line) Average shear
stress at the inner cylinder $S$ as a function of the confining
pressure $P$ for different cylinder roughness: ($\blacksquare$,
$\square$) sandpaper, ($\textcolor[rgb]{0.00,0.00,1.00}{\bullet}$)
glued seeds and
($\textcolor[rgb]{0.98,0.00,0.00}{\blacktriangle}$,
$\textcolor[rgb]{0.98,0.00,0.00}{\vartriangle}$) corrugated. The
full symbols indicate smooth horizontal walls, while hollow ones
indicate rough horizontal walls ($V_{\theta}=3.14~mm/s$)}}
\end{center}
\end{figure}

The average shear stress at the inner cylinder is simply deduced
from the measured torque $C$ as $S=C/(2\pi R_{i}^2H)$.
Consistently with previous studies
\cite{Coste04,Lerat96,Uesugi86a,Uesugi86b}, a linear relation
between the shear stress $S$ and the confining pressure $P$ is
observed (Fig.~\ref{Fig10}), from which an effective friction
coefficient $S/P$ is deduced (considering that the normal stress
at the cylinder wall is approximately equal to $P$) for each of
the three cylinders: $0.35$ for sandpaper, $0.38$ for glued seeds
and $0.4$ for corrugation. This increase seems consistent with the
decrease of the particle sliding shown in Fig.~\ref{Fig9}. The
roughness of the horizontal walls does not affect this value,
which is consistent with the very localized influence of this
roughness.

\subsection{Solid fraction profile}\label{solidprofiles}

\begin{figure}[!htb]
\begin{center}
\includegraphics*[width=6cm]{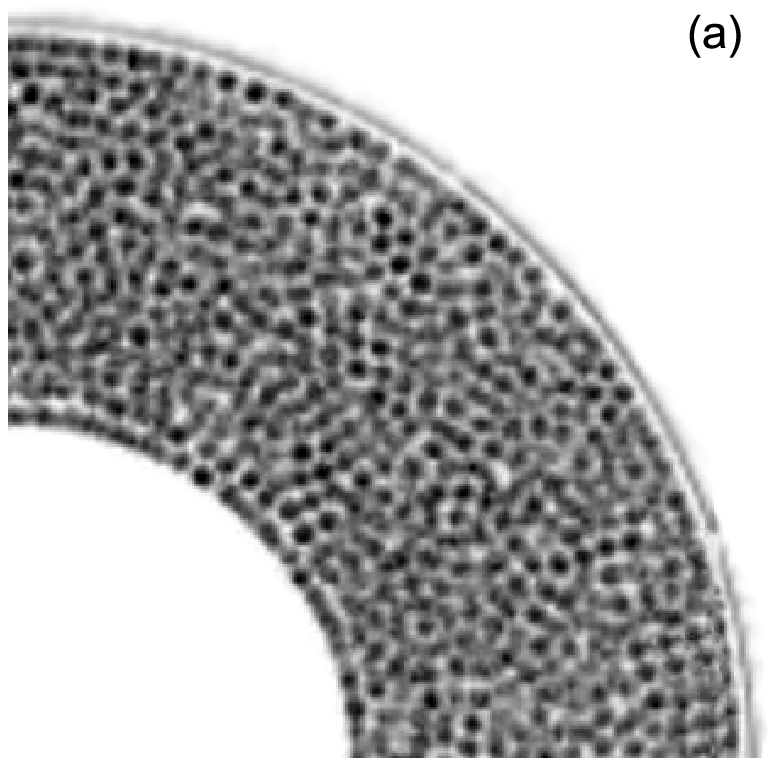}
\includegraphics*[width=6cm]{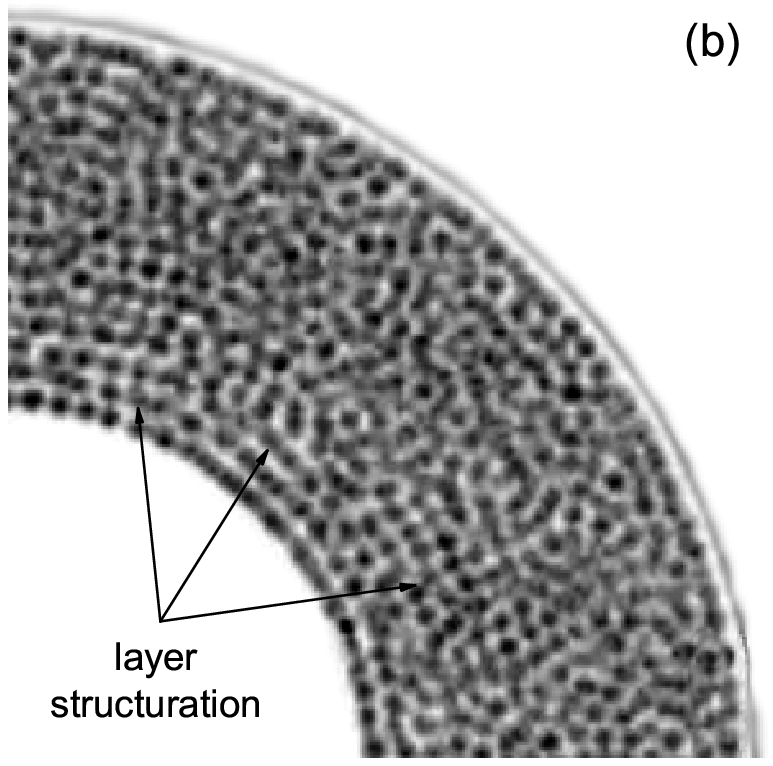}
\caption{\label{Fig11} \textit{Static images used for the
measurement of the solid fraction profile (sample quarter).
Definition: $25~pixels/mm^2$. (a) Before and (b) after shear
(corrugated cylinder).}}
\end{center}
\end{figure}

In order to measure the influence of the shear on the radial
distribution of grains, $3D$ spin density pictures ($20$ layers
$5~mm$ thick in $z$ direction, with a space resolution of $0.2~mm$
in both $x$ and $y$ directions) were performed at rest before and
after shearing (at least $200$ rotations, making $\approx 38~m$ of
tangential displacement). Fig.~\ref{Fig11}a and Fig.~\ref{Fig11}b
show typical density pictures obtained before and after shearing
respectively.

\begin{figure}[!htb]
\begin{center}
\includegraphics*[width=7.5cm]{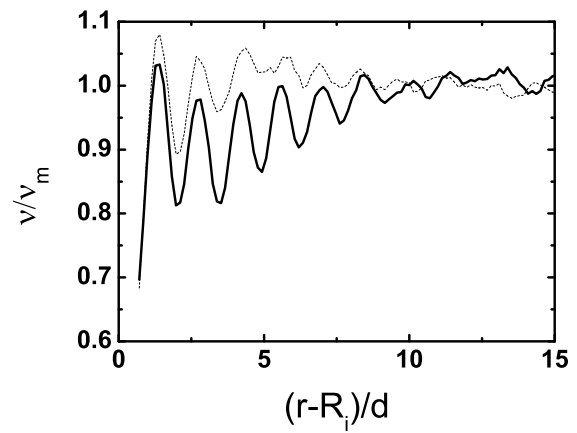}
\caption{\label{Fig12} \textit{Normalized solid fraction profiles
$\nu/\nu_m$ before shear (dotted line) and after shear (continuous
line).}}
\end{center}
\end{figure}

For each radial coordinate, grayscale pictures were then
integrated together in the orthoradial direction. Granted that the
integrated spin density profile in radial direction provides a
relative measurement of the solid fraction, these solid fractions
$\nu(r)$ are normalized by the mean value $\nu_m$ obtained before
shearing, and shown in Fig.~\ref{Fig12}.

Before shearing, after the granular material has been poured in
the cell, the solid fraction is relatively homogeneous, with a
certain structuration ($\nu(r)$ oscillations) near the inner wall.
After shearing, a much clearer structuration (already visible on
the static images (Fig.~\ref{Fig11})) together with a significant
dilation is observed close to the shearing cylinder
($r-R_{i}\lessapprox 7d $). However, the zone $7d \lessapprox
r-R_{i} \lessapprox 12d$ does not reveal significant solid
fraction changes. This indicates that most of the volumetric
variations are localized in the region near the cylinder, as
reported by other experiments and
simulations~\cite{Koval08,Koval09b,Koval09a}.

\section{Conclusions}\label{conclusions}

As a way to study the rheology of granular materials near an
interface, we have developed a cylinder shear apparatus inserted
in the MRI device, which allows the control of the radial
confining pressure exerted by the outer wall on the grains and the
measurement of the torque on the inner shearing cylinder. We have
also developed a multi-layer velocimetry sequence for the
simultaneous measurement of velocity profiles in different sample
zones. This study shows that the roughness of the shearing
interface significantly affects the granular interface rheology
(sliding velocity and friction coefficient), while the influence
of the roughness of the transverse confining walls remains
localized in the very first layers. Moreover the measurement of
the solid fraction profiles, based on static imaging of the
sample, shows that, when starting to shear the material, its
volumetric variations remains localized close to the shearing
interface. Those data are useful to test models attempting to
describe granular interface rheology~\cite{Mohan02, Latzel03,
Artoni08} for annular shear. Using this device together with the
velocimetry sequence, work is under progress to study the
transient behavior of the granular material, under monotonic or
cyclic shear, and the influence of liquid saturation.


\begin{thebibliography}{10}

\bibitem{Nedderman92}
R.M. Nedderman.
\newblock {\em Statics and kinematics of granular materials}.
\newblock Cambridge University Press, Cambridge, 1992.

\bibitem{Schlosser78}
F.~Schlosser and V.~Elias.
\newblock Friction in reinforced earth.
\newblock In {\em Proceedings of the International Symposium on Earth
  Reinforcement}, pages 735--763, Pittsburgh, Pennsylvania, April 1978.

\bibitem{Tubacanon95}
J.T. Tubacanon, D.W. Airey, and H.G. Poulos.
\newblock Pile skin friction in sands from constant normal stiffness tests.
\newblock {\em Geotechnical Testing Journal}, 18(3):350--364, 1995.

\bibitem{Chambon06}
G.~{Chambon}, J.~{Schmittbuhl}, and A.~{Corfdir}.
\newblock Frictional response of a thick gouge sample: 2. friction law and
  implications for faults.
\newblock {\em J. Geophys. Res.}, 111:B09309, 2006.

\bibitem{Chevoir09}
F.~Chevoir.
\newblock {\em Granular flows (in French)}.
\newblock Laboratoire Central des Ponts et Chauss\'ees - Collection Etudes et
  Recherches des Laboratoires des Ponts et Chauss\'ees, 2009.

\bibitem{DeJong03}
J.T. DeJong, M.F. Randolph, and D.J. White.
\newblock Interface load transfer degradation during cyclic loading: a
  microscale investigation.
\newblock {\em Soils and Foundations}, 43(4):81--93, 2003.

\bibitem{Koval08}
G.~Koval.
\newblock {\em Interface behavior of granular materials (in French)}.
\newblock PhD thesis, École des Ponts Paristech, Paris, 2008.
\newblock http://tel.archives-ouvertes.fr/tel-00311984.

\bibitem{Oumarou05}
T.A. Oumarou and E.~Evgin.
\newblock Cyclic behaviour of a sand-steel plate interface.
\newblock {\em Can. Geotech. J.}, 42:1695--1704, 2005.

\bibitem{Uesugi86a}
M.~Uesugi and H.~Kishida.
\newblock Influential factors of friction between steel and dry sands.
\newblock {\em Soils and Foundations}, 26:33--46, 1986.

\bibitem{Uesugi86b}
M.~Uesugi and H.~Kishida.
\newblock Frictional resistance at yield between dry sands and steel.
\newblock {\em Soils and Foundations}, 26:139--149, 1986.

\bibitem{Uesugi88}
M.~Uesugi, H.~Kishida, and Y.~Tsubakihara.
\newblock Behavior of sand particules in sand-steel friction.
\newblock {\em Soils and foundations}, 28(1):107--118, 1988.

\bibitem{Vardoulakis95}
I.~Vardoulakis and P.~Unterreiner.
\newblock Interfacial localisation in simple shear tests on a granular medium
  modelled as a cosserat continuum.
\newblock In A.P.S. Selvadurai and M.~Boulon, editors, {\em Mechanics of
  Geomaterial Interfaces}, volume~42, pages 487--512, 1995.

\bibitem{Koval09a}
G.~Koval, J-N. Roux, A.~Corfdir, and F.~Chevoir.
\newblock Annular shear of cohesionless granular materials: From the inertial
  to quasistatic regime.
\newblock {\em Phys. Rev. E}, 79:021306, 2009.

\bibitem{Howell99a}
D.~Howell, R.P. Behringer, and C.T. Veje.
\newblock Stress fluctuations in a {2D} granular {C}ouette experiment : a
  continuum transition.
\newblock {\em Phys. Rev. Lett.}, 82:5241--5244, 1999.

\bibitem{Latzel03}
M.~Latzel, S.~Luding, H.J. Herrmann, D.W. Howell, and R.P.
Behringer.
\newblock Comparing simulation and experiment of a {2D} granular {C}ouette
  shear device.
\newblock {\em Euro. Phys. J. E}, 11:325--333, 2003.

\bibitem{Miller96}
B.~Miller, C.~O'Hern, and R.~P. Behringer.
\newblock Stress fluctuations for continuously sheared granular materials.
\newblock {\em Phys. Rev. Lett.}, 77:3110--3113, 1996.

\bibitem{Veje99}
C.T. Veje, D.W. Howell, and R.P. Behringer.
\newblock Kinematics of two - dimensional granular {C}ouette experiment at the
  transition to shearing.
\newblock {\em Phys. Rev. E}, 59:739--745, 1999.

\bibitem{Bocquet02b}
L.~{Bocquet}, W.~{Losert}, D.~{Schalk}, T.~C. {Lubensky}, and
J.~P. {Gollub}.
\newblock Granular shear flow dynamics and forces : Experiment and continuous
  theory.
\newblock {\em Phys. Rev. E}, 65:011307, 2002.

\bibitem{Chambon03}
G.~{Chambon}, J.~{Schmittbuhl}, A.~{Corfdir}, J.-P. {Vilotte}, and
S.~{Roux}.
\newblock Shear with comminution of a granular material: Microscopic
  deformations outside the shear band.
\newblock {\em Phys. Rev. E}, 68:011304, 2003.

\bibitem{Dacruz04a}
F.~da~Cruz.
\newblock {\em Flow of dry grains : Friction and jamming (in French)}.
\newblock PhD thesis, École des Ponts Paristech, Paris, 2004.
\newblock http://pastel.paristech.org/archive/00000946.

\bibitem{Dacruz02}
F.~da~Cruz, F.~Chevoir, D.~Bonn, and P.~Coussot.
\newblock Viscosity bifurcation in granular materials, foams, and emulsions.
\newblock {\em Phys. Rev. E}, 66:051305, 2002.

\bibitem{Daniel07}
R.C. Daniel, A.P. Poloski, and A.E. Saez.
\newblock Vane rheology of cohesionless glass beads.
\newblock {\em Powder Tech.}, 179:62--73, 2007.

\bibitem{Losert01}
W.~Losert and G.~Kwon.
\newblock Transition and steady-state dynamics of granular shear flows.
\newblock {\em Advances in complex systems}, 4:369--377, 2001.

\bibitem{Mueth03}
D.M. {Mueth}.
\newblock Mesurement of particule dynamics in slow, dense granular {C}ouette
  flow.
\newblock {\em Phys. Rev. E}, 67:011304, 2003.

\bibitem{Mueth00}
D.M. Mueth, G.F. Debregeas, G.S. Karczmar, P.J. Eng, S.R. Nagel,
and H.M.
  Jaeger.
\newblock Signature of granular microstructure in dense shear flows.
\newblock {\em Nature}, 406:385--389, 2000.

\bibitem{Tardos98}
G.I. {Tardos}, M.I. {Khan}, and D.G. {Schaeffer}.
\newblock Forces on a slowly rotating, rough cylinder in a {C}ouette device
  containing a dry, frictional powder.
\newblock {\em Phys. Fluids}, 10:335--341, 1998.

\bibitem{Tardos03}
G.I. Tardos, S.~McNamara, and I.~Talu.
\newblock Slow and intermediate flow of a frictional bulk powder in the
  {C}ouette geometry.
\newblock {\em Powder Tech.}, 131:23--39, 2003.

\bibitem{Wang08}
P.~Wang, C.~Song, C.~Briscoe, and H.A. Makse.
\newblock Particle dynamics and effective temperature of jammed granular matter
  in a slowly sheared {3D} {C}ouette cell.
\newblock {\em Phys. Rev. E}, 77:061309, 2008.

\bibitem{Koval09b}
G.~{Koval}, F.~{Chevoir}, J.-N. {Roux}, J.~{Sulem}, and
A.~{Corfdir}.
\newblock Slow annular shear of granular-continuum interfaces: macroscopic and
  mesoscopic observations.
\newblock submitted to Granular Matter, 2009.

\bibitem{Fukushima99}
E.~Fukushima.
\newblock Nuclear magnetic resonance as a tool to study flow.
\newblock {\em Annu. Rev. Fluid Mech.}, 31:95--123, 1999.

\bibitem{Fukushima06}
E.~Fukushima.
\newblock {\em NMR Imaging in Chemical Engineering}, chapter Granular flow,
  pages 490--508.
\newblock Wiley-VCH, 2006.

\bibitem{Kuperman96}
V.~Y. {Kuperman}.
\newblock Nuclear magnetic resonance measurements of diffusion in granular
  media.
\newblock {\em Phys. Rev. Lett.}, 77:1178--1181, 1996.

\bibitem{Nakagawa93}
M.~Nakagawa, S.A. Altobelli, A.~Caprihan, E.~Fukushima, and E.K.
Jeong.
\newblock Non-invasive measurements of granular flow by magnetic resonance
  imaging.
\newblock {\em Exp. in Fluids}, 16:54--60, 1993.

\bibitem{Caprihan00}
A.~{Caprihan} and J.D. {Seymour}.
\newblock Correlation time and diffusion coefficient imaging : application to
  granular flow system.
\newblock {\em Journal of Magnetic Resonance}, 144:96--107, 2000.

\bibitem{Ristow99}
G.H. Ristow and M.~Nakagawa.
\newblock Shape dynamics of interfacial front in rotating cylinders.
\newblock {\em Phys. Rev. E}, 59:2044--2048, 1999.

\bibitem{Seymour00}
J.~D. Seymour, A.~Caprihan, S.~Altobelli, and E.~Fukushima.
\newblock Pulsed gradient spin echo nuclear magnetic resonance imaging of
  diffusion in granular flow.
\newblock {\em Phys. Rev. Lett.}, 84:266--269, 2000.

\bibitem{Yamane98}
K.~Yamane, M.~Nakagawa, S.~A. Altobelli, T.~Tanaka, and Y.~Tsuji.
\newblock Steady particulate flows in a horizontal rotating cylinder.
\newblock {\em Phys. Fluids}, 10:1419--1427, 1998.

\bibitem{Chevoir01b}
F.~Chevoir, M.~Prochnow, P.~Moucheront, F.~da~Cruz, F.~Bertrand,
J.P. Guilbaud,
  P.~Coussot, and J.N. Roux.
\newblock Dense granular flows in a vertical chute.
\newblock In Y.~Kishino, editor, {\em Powders and grains 2001}, pages 399--402,
  Rotterdam, 2001. Balkema.

\bibitem{Cheng06}
X.~Cheng, J.~B. Lechman, A.~Fernandez-Barbero, G.~S. Grest, H.~M.
Jaeger, G.~S.
  Karczmar, M.~E. M\"{o}bius, and S.~R. Nagel.
\newblock Three-dimensional shear in granular flow.
\newblock {\em Phys. Rev. Lett.}, 96:038001, 2006.

\bibitem{Caprihan97}
A.~Caprihan, E.~Fukushima, A.~D. Rosato, and M.~Kos.
\newblock Magnetic resonance imaging of vibrating granular beds by spatial
  scanning.
\newblock {\em Rev. Sci. Instrum.}, 68:4217--–4220, 1997.

\bibitem{Ehrichs95}
E.~E. {Ehrichs}, H.~M. {Jaeger}, G.~S. {Karczmar}, J.~B. {Knight},
V.~Y.
  {Kuperman}, and S.~R. {Nagel}.
\newblock Granular convection observed by magnetic resonance imaging.
\newblock {\em Science}, 267:1632--1634, 1995.

\bibitem{Hill97c}
K.~M. Hill, A.~Caprihan, and J.~Kakalios.
\newblock Bulk segregation in rotated granular material measured by magnetic
  resonance imaging.
\newblock {\em Phys. Rev. Lett.}, 78:50--53, 1997.

\bibitem{Knight96}
J.~B. Knight, E.~E. Ehrichs, V.~Y. Kuperman, J.~K. Flint, H.~M.
Jaeger, and
  S.~R. Nagel.
\newblock Experimental study of granular convection.
\newblock {\em Phys. Rev. E}, 54:5726--5738, 1996.

\bibitem{Kuperman95}
V.~Y. Kuperman, E.~E. Ehrichs, H.~M. Jaeger, and G.~S. Karczmar.
\newblock A new technique for differentiating between diffusion and flow in
  granular media using magnetic resonance imaging.
\newblock {\em Rev. Sci. Instrum.}, 66:4350--4355, 1995.

\bibitem{Metcalfe96}
G.~Metcalfe and M.~Shattuck.
\newblock Pattern formation during mixing and segregation of flowing granular
  materials.
\newblock {\em Physica A}, 233:709 -- 717, 1996.

\bibitem{Yang00a}
X.~Yang and D.~Candela.
\newblock Potential energy in a three-dimensional vibrated granular medium
  measured by {NMR} imaging.
\newblock {\em Phys. Rev. Lett.}, 85:298--301, 2000.

\bibitem{Yang02}
X.~Yang, Y.~Huan, D.~Candela, R.W. Mair, and R.L. Walsworth.
\newblock Measurement of grain motion in a dense, three-dimensional granular
  fluid.
\newblock {\em Phys. Rev. Lett.}, 88:044301, 2002.

\bibitem{Corfdir04}
A.~Corfdir, P.~Lerat, and I.~Vardoulakis.
\newblock A cylinder shear apparatus.
\newblock {\em Geotechnical Testing Journal}, 27:447--455, 2004.

\bibitem{Dumitrescu05}
A.I. Dumitrescu.
\newblock {\em Experimental study of the interface behavior between a structure
  and a granular soil (in French)}.
\newblock PhD thesis, École des Ponts Paristech, Paris, 2005.

\bibitem{Lerat96}
P.~Lerat.
\newblock {\em Study of the soil - structure interface in granular media by
  using a new ring shear apparatus (in French)}.
\newblock PhD thesis, École des Ponts Paristech, Paris, 1996.

\bibitem{Lerat97}
P.~Lerat, M.~Boulon, and F.~Schlosser.
\newblock Experimental study of the soil - structure interface in granular
  media.
\newblock {\em Revue Française de Génie Civil}, 1:345--366, 1997.

\bibitem{Hanlon98}
A.D. Hanlon, S.J. Gibbs, L.D. Hall, D.E. Haycock, W.J. Frith, and
S.~Ablett.
\newblock Rapid {MRI} and velocimetry of cylindrical {C}ouette flow.
\newblock {\em Magn. Reson. Imaging}, 16:953–961, 1998.

\bibitem{Bonn08}
D.~Bonn, S.~Rodts, M.~Groenink, S.~Rafaï, N.~Shahidzadeh-Bonn, and
P.~Coussot.
\newblock Some applications of magnetic resonance imaging in fluid mechanics:
  Complex flows and complex fluids.
\newblock {\em Annu. Rev. Fluid Mech.}, 40:209–233, 2008.

\bibitem{Huang05}
N.~{Huang}, G.~{Ovarlez}, F.~{Bertrand}, S.~{Rodts}, P.~{Coussot},
and
  D.~{Bonn}.
\newblock Flow of wet granular materials.
\newblock {\em Phys. Rev. Lett.}, 94:028301, 2005.

\bibitem{Jarny05}
S.~Jarny, N.~Roussel, S.~Rodts, F.~Bertrand, R.~Le~Roy, and
P.~Coussot.
\newblock Rheological behavior of cement pastes from {MRI} velocimetry.
\newblock {\em Cement and Concrete Research}, 35:1873--1881, 2005.

\bibitem{Ovarlez06}
G.~Ovarlez, F.~Bertrand, and S.~Rodts.
\newblock Local determination of the constitutive law of a dense suspension on
  noncolloidal particles through magnetic resonance imaging.
\newblock {\em J. Rheol.}, 50:256--292, 2006.

\bibitem{Ovarlez08}
G.~Ovarlez, S.~Rodts, A.~Ragouilliaux, P.~Coussot, J.~Goyon, and
A.~Colin.
\newblock Wide-gap couette flows of dense emulsions: Local concentration
  measurements, and comparison between macroscopic and local constitutive law
  measurements through magnetic resonance imaging.
\newblock {\em Phys. Rev. E}, 78:036307, 2008.

\bibitem{Moller08}
P.~C. F.~M\o ller, S.~Rodts, M.~A.~J. Michels, and D.~Bonn.
\newblock Shear banding and yield stress in soft glassy materials.
\newblock {\em Phys. Rev. E}, 77:041507, 2008.

\bibitem{Raynaud02}
J.S. Raynaud, P.~Moucheront, J.C. Baudez, F.~Bertrand, J.P.
Guilbaud, and
  P.~Coussot.
\newblock Direct determination by nmr of the thixotropic and yielding behavior
  of suspensions.
\newblock {\em J. Rheol.}, 46:709--732, 2002.

\bibitem{Rodts04}
S.~Rodts, F.~Bertrand, S.~Jarny, P.~Poullain, and P.~Moucheront.
\newblock Recent developments in {MRI} applications to rheology and fluid
  mechanics.
\newblock {\em Comptes Rendus Chimie}, 7:275 --282, 2004.

\bibitem{Jop05b}
P.~Jop, Y.~Forterre, and O.~Pouliquen.
\newblock Crucial role of sidewalls in granular surface flows: consequences for
  the rheology.
\newblock 541:167--192, 2005.

\bibitem{Jop06}
P.~Jop, Y.~Forterre, and O.~Pouliquen.
\newblock A constitutive law for dense granular flows.
\newblock {\em Nature}, 441:727--730, 2006.

\bibitem{Coste04}
C.~{Coste}.
\newblock {Shearing of a confined granular layer: Tangential stress and
  dilatancy}.
\newblock {\em Phys. Rev. E}, 70:051302, 2004.

\bibitem{Mohan02}
L.~S. {Mohan}, K.~K. {Rao}, and P.~R. {Nott}.
\newblock Frictional {C}osserat model for slow shearing of granular materials.
\newblock 457:377--409, 2002.

\bibitem{Artoni08}
R.~Artoni, P.~Canu, and A.~Santomaso.
\newblock Effective boundary conditions for dense granular flows.
\newblock {\em Phys. Rev. E}, 79:031304, 2009.

\end{thebibliography}

\end{document}